# Superconducting-insulating phase transition in pressurized $Ba_{1-x}K_xBiO_3$

Jinyu Han[1,6] *, Xiangde Zhu[2]*, Jianfeng Zhang[1]*, Shu Cai[3]*, Luhong Wang[4]*, Yang Gao[3], Fuyang Liu[3], Haozhe Liu[3], Saori I. Kawaguchi[5], Jing Guo[1], Yazhou Zhou[1], Jinyu Zhao[1,6], Pengyu Wang[1,6], Lixin Cao[1], Mingliang Tian[2], Qi Wu[1], Tao Xiang[1,6,7] and Liling Sun[1,3,6,†]

*[1]Institute of Physics, Chinese Academy of Sciences, Beijing 100190, China*

*[2] Anhui Key Laboratory of Low Energy Quantum Materials and Devices, High Magnetic Field Laboratory, Chinese Academy of Sciences, Hefei, Anhui 230031, China*

*[3] Center for High Pressure Science & Technology Advanced Research, 100094 Beijing, China*

*[4]Shanghai Advanced Research in Physical Sciences, Shanghai 201203, China*

*[5]Japan Synchrotron Radiation Research Institute, SPring-8, Sayo-gun Hyogo 679-5198, Japan*

*[6]University of Chinese Academy of Sciences, Beijing 100190, China*

*[9]Beijing Academy of Quantum Information Sciences, Beijing 100193, China*

We report the first observation of a pressure-induced transition from a superconducting (SC) to an insulating (I) phase in single-crystal $Ba_{1-x}K_xBiO_3$ (x = 0.4, 0.43, 0.52, and 0.58) superconductors. X-ray diffraction measurements conducted at 20 K reveal a direct relationship between this SC-I transition and a pressure-induced distortion of crystal structure. With increasing pressure, the lattice parameters *a* and *c* of the ambient-pressure superconducting tetragonal (T) phase are compressed continuously below a critical pressure ($P_{c1}$), wherein the pressure ($P$) dependence of superconducting transition temperature ($T_c$) displays a small variation. However, upon further compression, the lattice of the compressed T phase displays an anisotropic change, and $T_c$ shows a monotonous decrease. When the pressure reaches $P_{c2}$ ($P_{c2} > P_{c1}$), the compressed T phase collapses along the *c* axis, followed by the disappearance of superconductivity and the appearance of the insulating phase. This SC-I transition is fully reversible, with the critical pressure increasing alongside K doping concentration. These findings are strikingly similar to the SC-I transition observed in hole-doped high-$T_c$ cuprate superconductors under pressure. Identifying their commonalities could deepen our understanding of the mechanisms that underlie high- $T_c$ superconductivity in these two oxide superconductors with a perovskite structure.

Soon after the discovery of high-temperature superconductors in perovskite copper-oxides (cuprates) [1,2], a new class of perovskite high-temperature superconductor was found in bismuth-oxide $Ba_{1-x}K_xBiO_3$, (bismuthates), sparking a renewed excitement in the scientific community [3-5]. There is a prevailing consensus that the superconductivity induced by doping in bismuthates is intimately associated with their electronic and crystal structures. At low temperature, the parent compound $BaBiO_3$ exhibits insulating properties with a commensurate charge-density-wave (CDW) order considered to arise from either a disproportionation of Bi cations of $Bi^{+3}$ and $Bi^{+5}$ [6-8] or a Bi–O charge-transfer mechanism leading to charge localization on oxygen orbitals [9,10]. Upon the potassium (K) doping, this modulated ordered phase disappears, giving rise to superconductivity [3,7,11,12]. As doping level reaches about 0.4, superconducting transition temperature ($T_c$) of the $Ba_{1-x}K_xBiO_3$ superconductors approaches its maximum [3]. Despite extensive efforts have been made to uncover further intriguing phenomena in the bismuthate superconductors, the method of chemical doping has yielded only limited results [11-15].

Pressure is an effective and clean method to tune the superconductivity and unveil novel physics by tunning the interactions among multiple degrees of freedom in materials without introducing chemical complexities. This technique has been widely utilized as an independent control parameter in the studies of superconducting and other corelated electron systems [16-35]. Compelling examples include the enhancement of superconducting transition temperature ($T_c$) under pressure in mercury-containing cuprates and Fe-pnictides [27,30]. Recently, an intriguing phenomenon of the pressure-induced superconducting (SC) phase to an insulating-like (I) phase was found in hole-doped cuprate superconductors [35], highlighting the significance of pressure as an effective tool for exploring new physics. However, a systematic investigation on the high-pressure transport and structural properties of the $Ba_{1-x}K_xBiO_3$ samples is lacking, except for the one reporting the studies on $Ba_{1-x}K_xBiO_3$ up to 8 GPa [36]. In this work, we present the first results of high-pressure properties measured from $Ba_{1-x}K_xBiO_3$ up to 44.6 GPa, which reveals the universal presence of a pressure-induced superconducting to insulating (SC-I) phase transition in this material. Our finding is supported by a combination of high-pressure resistance and low-temperature synchrotron X-ray diffraction measurements.

High-quality single crystals of $Ba_{1-x}K_xBiO_3$ were grown using a modified electrochemical method [37]. High-pressure resistance measurements were performed in a diamond-anvil cell (DAC), in which diamond anvils with 300 μm and 400 μm culets (flat area of the diamond anvil) were employed for independent measurements. Au foil was used as electrodes, rhenium plate with 100-μm-diameter hole as a gasket and cubic boron nitride as an insulating material. The standard four-probe electrodes were applied in the resistance measurements on the $Ba_{1-x}K_xBiO_3$ single crystals. To provide a quasi-hydrostatic pressure environment for the sample, NaCl powder was employed as the pressure transmitting medium. The pressure of the sample was determined using the ruby fluorescence method [43]. High-pressure and low-temperature X-ray diffraction (XRD) measurements were conducted at Spring-8 in Japan. A monochromatic X-ray beam with a wavelength of 0.4138 Å was employed for the measurements. A diamond anvil cell (DAC) with the loaded sample and the gas membrane system was placed in a cryostat at room temperature and then was cooled from 300 K down to 20 K. Subsequently, we increased pressure at 20 K and collected XRD patterns at each pressure point. In the high-pressure XRD measurements, elemental Au was used as a pressure standard to determine the pressure of the sample [44-46].

We first performed single crystal XRD diffraction measurements on one of the superconducting samples ($Ba_{0.6}K_{0.4}BiO_3$) at ambient-pressure and confirmed that it crystallizes in a cubic unit cell at room temperature, with lattice parameters $a$ = 4.34 Å and space group *Pm-3m* (No.221), as shown in Fig.1a-1d. Our characterized result is in good agreement with the result reported previously [11]. Since $Ba_{1-x}K_xBiO_3$ is sensitive to air and moisture, we loaded the sample into diamond anvil cell (DAC) in a glovebox for each run, and subsequently transferred the DAC with the loaded sample from the glovebox to the cryostat for the near-ambient-pressure resistance measurement. As shown in Fig.1e, the plot of resistance versus temperature measured at 1.2 GPa shows a sudden drop at ~ 31 K, followed by reaching a zero-resistance state. When a magnetic field is applied, the resistance drop shifts to a lower temperature, implying a superconducting transition. The superconducting transition was further confirmed by the measurements of the magnetic moment in zero-field cooling (ZFC) and field-cooling (FC) modes, which show diamagnetic throws at ~ 31 K (Fig.1f). These results align with the results reported previously [11,14].

Next, we conducted high-pressure resistance measurements on the superconducting $Ba_{1-x}K_xBiO_3$ samples with different doping levels (Fig.2). The results of resistance versus temperature obtained at different pressures show that the $T_c$ values of the $x$=0.40, 0.43 0.52 and 0.58 samples are 31 K at 1.2 GPa (Fig.2a), 28.5 K at 1.1 GPa (Fig. 2b), 18.1 K at 1.3 GPa (Fig.2c) and 14.1 K at 1.4 GPa (Fig.2d), respectively. The pressure-$T_c$ phase diagrams for the measured samples are shown in Fig.3a-3e. The results exhibit the same high-pressure behavior: $T_c$ displays slight variation below the first critical pressure ($P_{c1}$) and then exhibits a monotonous decrease with increasing pressure up to the second critical pressure ($P_{c2}$), at which the superconducting state is fully suppressed, and an insulating state appears. We repeated the measurements on new samples and found that the results were reproducible [37]. Significantly, we noticed that a higher doping level in the sample requires an increased critical pressure to trigger the transition from the superconducting (SC) to the insulating (I) phase (Fig. 3f). This is reminiscent what we have seen in $Bi_{2-x}Sr_xCaCu_2O_{8+\delta}$ [35], where a similar behavior is observed. To visualize the high-pressure behavior in these two superconductors, we include the results of hole doping concentration ($p$) per Cu versus the critical pressure for the SC-I phase transition ($P_{c2}$) for $Bi_2Sr_2CaCu_2O_{8+\delta}$ in Fig.3f. It is found that they have the same trend.

To know whether this SC-I phase transition of the investigated samples is associated with the pressure-induced crystal structural phase transition, we performed high-pressure and low-temperature synchrotron X-ray diffraction measurements on the $x$=0.40 sample at SPring 8 in Japan (Fig.4). Given that the low-temperature crystal structure of the superconducting $Ba_{1-x}K_xBiO_3$ crystalizes in the tetragonal *I₄/mam* (T) unit at low temperature and ambient pressure [12], it is important to trace the effect of pressure on its structural stability under low temperature condition. At 2.2 GPa, we found that the sample also crystallizes in the T phase with space group *I4/mcm* (Fig.4a), in good agreement with the results reported previously [12]. Subsequently, we maintained the sample at the fixed temperature of 20 K and increased the pressure up to 32.1 GPa, which exceeds the pressure required for the SC-I phase transition (see Fig.4b). Although we did not observe the splits of higher angle (2θ> 55 degrees) peaks reported in Ref. [12], due to the constraints of the limited window opening of the DAC (the maximum 2θ is only about 25 degrees), however, we found splits of the low angle peaks starting at 11.7 GPa (see Fig.4c and 4d). The Rietveld structural refinements up to 32.1 GPa indicate that the sample still crystalizes in the tetragonal (T) unit cell in space group *I4/mcm* (Fig.4c-4d). The pressure dependence of the lattice parameter *a* and *c* will discuss below.

Then we combined our XRD results with the pressure-$T_c$ phase diagram in Fig.5. Structurally, the phase diagram can be divided into three distinct regions: the region of the tetragonal (T) phase (left), the distorted T phases (the middle) and the region of collapsed T phase (right). Below the critical pressure ($P_{c1}$) of 11.7 GPa, $T_c$ of the tetragonal $Ba_{0.6}K_{0.4}BiO_3$ remains relatively stable, suggesting that the pressure can maintain a balance between the crystal and electronic structures within this interval. In the pressure range between $P_{c1}$ and $P_{c2}$, the lattice of the compressed T phase displays an anisotropic change, the lattice parameter *a* shrinks slowly and $T_c$ shows a small variation. When the pressure reaches $P_{c2}$ (23.2 GPa), the T phase collapses along the *c* axis, followed by the disappearance of superconductivity and the appearance of the insulating phase. These results implies that the pressure-induced collapse of the compressed T phase alters the electronic structure that is crucial for the SC-I phase transition.

This pressure-induced SC-I phase transition observed in the $Ba_{1-x}K_xBiO_3$ superconductors is reminiscent what has been found in the pressurized hole-doped high-$T_c$ copper-oxide superconductors that also possess a perovskite structure [35], in which a pressure-induced SC-I phase transition is observed as well. It is notable that the crystal structure of the cuprates is found to be stable in the orthorhombic phase throughout the entire pressure range [35], while $Ba_{1-x}K_xBiO_3$ crystallizes in the T phase and displays a distinct collapse along the *c* axis at $P_{c2}$, implying that the origin driven the SC-I phase transition should be different in these two oxide superconductors with a perovskite structure. The comparative analysis of the pressure-induced SC-I phase transition in the bismuthate and cuprate superconductors is expected to provide new perspectives for a more profound understanding of high-$T_c$ superconductivity emerging from hole-doped metal-oxides with a perovskite structure.

**Acknowledgements**

We thank Prof. J. Zaanen for helpful discussions on this work. The work was supported by the National Key Research and Development Program of China (Grant No. 2022YFA1403900 and 2021YFA1401800), the NSF of China (Grants No. U2032214, 12104487, 12122414 and 12004419). The low-temperature XRD measurements were conducted at SPring-8 (Grants No. 2023B1182 and 2023B1273). J. G. and S.C. are grateful for supports from the Youth Innovation Promotion Association of the CAS (2019008) and the China Postdoctoral Science Foundation (E0BK111).

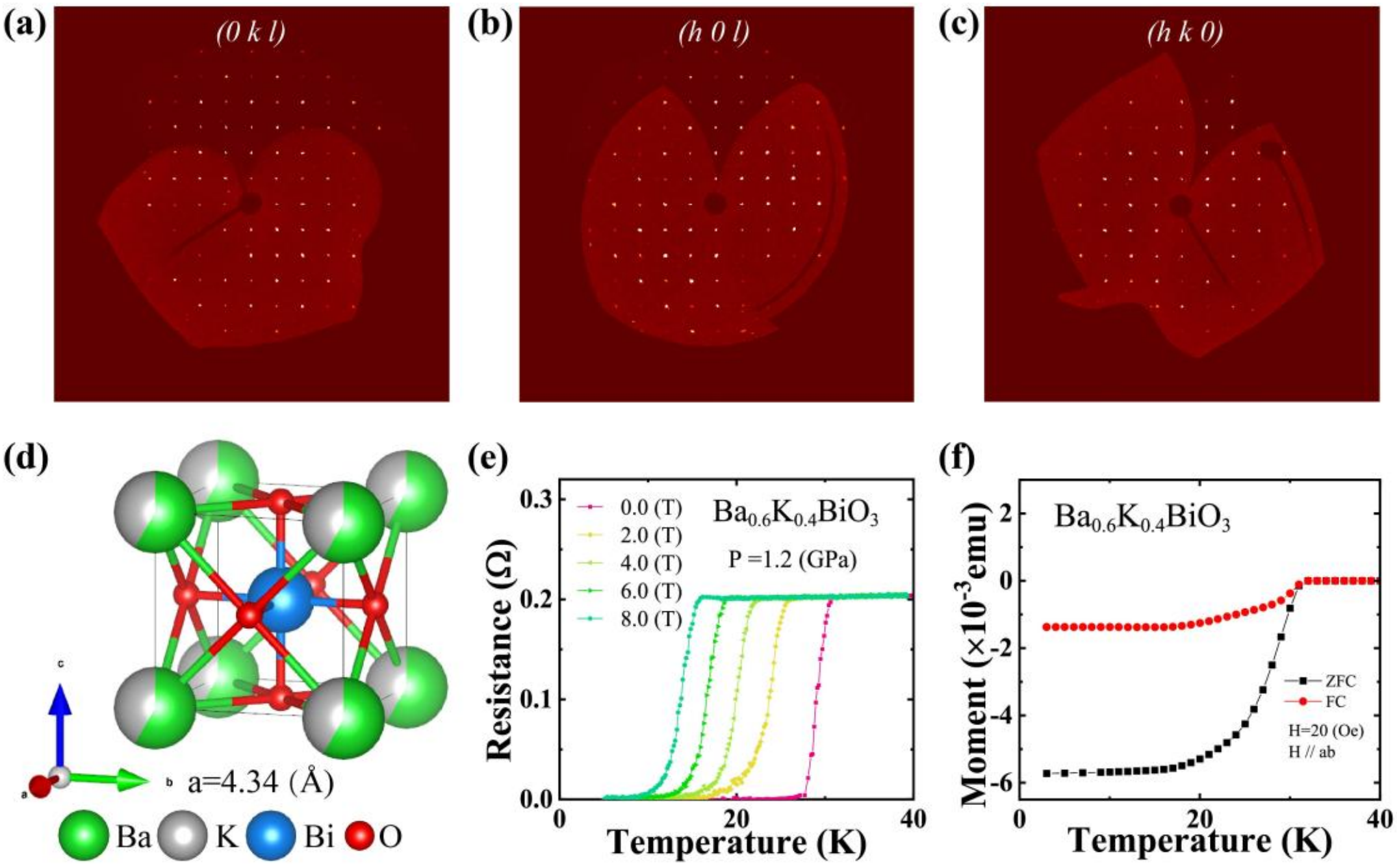

Figure 1 Structure characterization and transport measurements on the perovskite superconducting $Ba_{0.6}K_{0.4}BiO_3$. (a)-(c) The single crystal X-ray diffraction patterns for the (0*kl*), (*h*0*l*) and (*hk*0) zones of the sample, respectively, measured at 300 K. (d) Schematic crystal structure of the sample, illustrating the refined crystal structure with a cubic unit cell. (e) Resistance as a function of temperature at 1.2 GPa under different magnetic fields. (f) Magnetic moment versus temperature in both field-cooled (FC) and zero-field-cooled (ZFC) modes at ambient pressure.

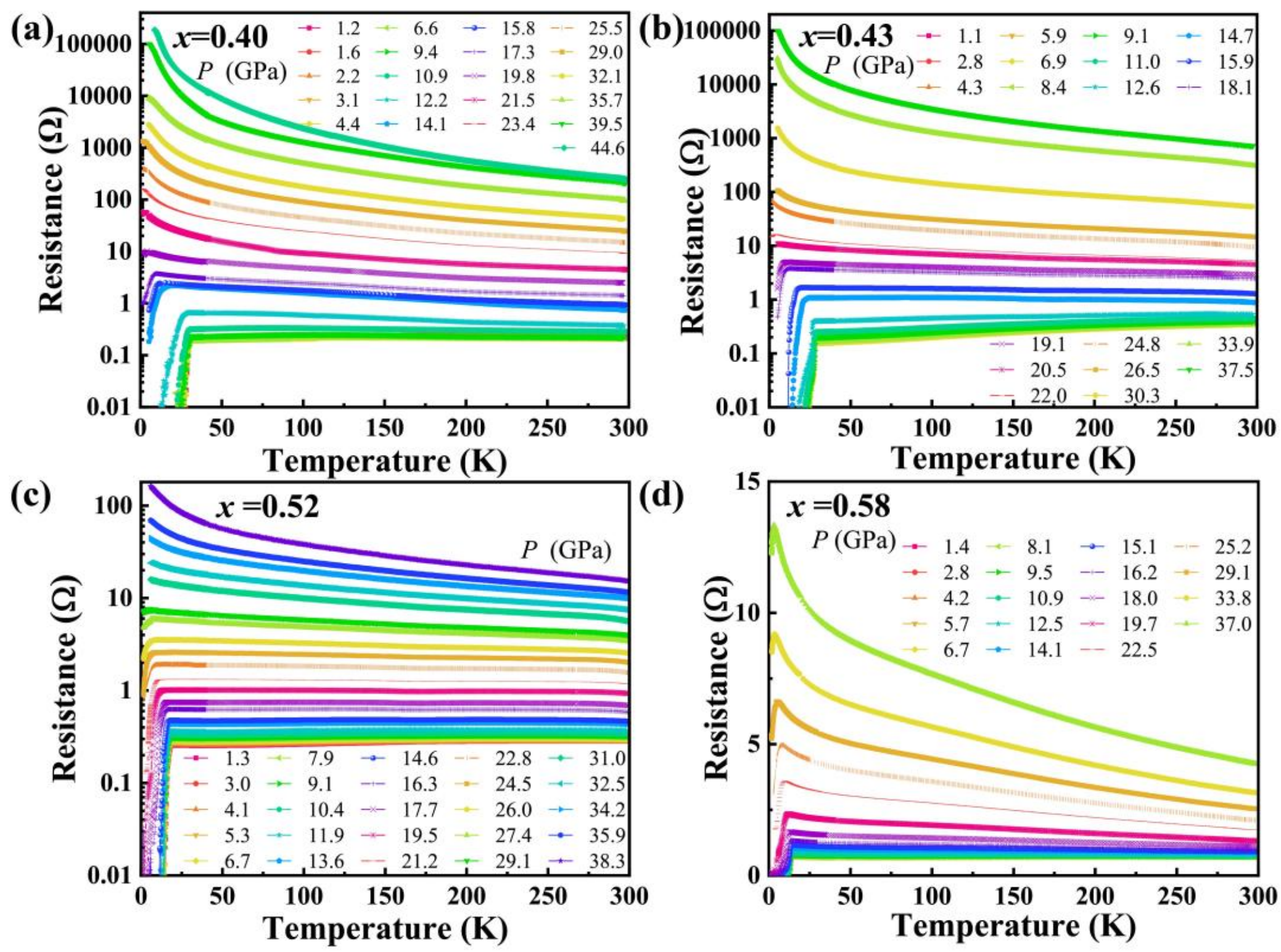

Figure 2 Temperature dependence of resistance for $Ba_{1-x}K_xBiO_3$ at different pressures. (a)The log-linear plots of resistance versus temperature for the $x$=0.40 sample measured in the pressure range 1.2 - 44.6 GPa. (b) for the $x$=0.43 sample measured in the pressure range 1.1 - 37.5 GPa. (c) for the $x$=0.52 sample measured in the pressure range 1.3 - 38.3 GPa. (d) for the $x$=0.58 sample measured in the pressure range 1.4 -37.0 GPa. All samples exhibit the same behavior - an insulating-like state appears above a critical pressure.

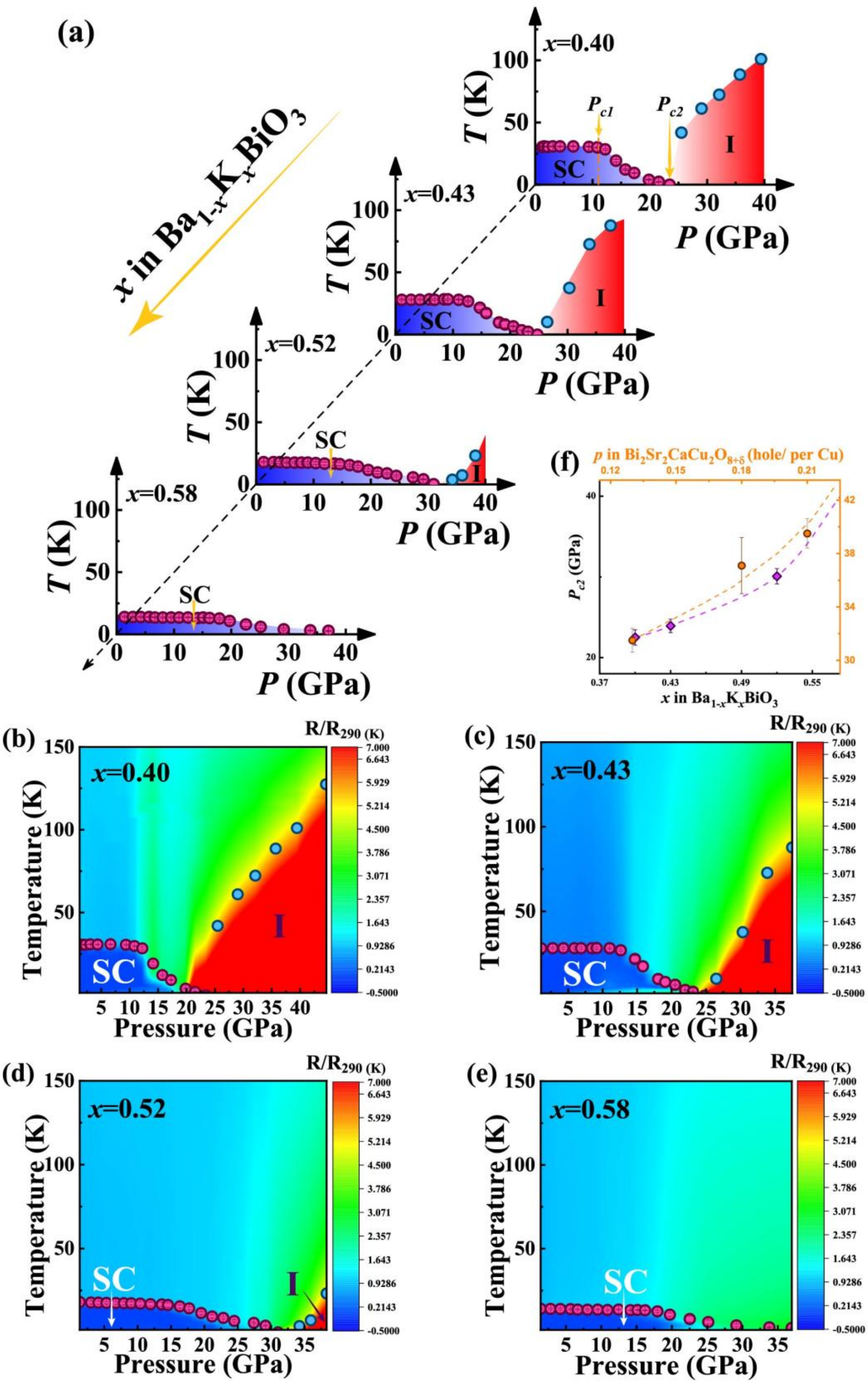

Figure 3 (a) Pressure - Temperature phase diagrams for the $Ba_{1-x}K_xBiO_3$

superconductors with different doping concentrations, which show a universal superconducting-insulating (SC-I) phase transition. Details regarding the temperature determination of the superconducting and insulating transitions can be found in Ref.37. It is seen that the higher the doping level, the higher the critical pressure required for the superconducting-insulating transition. $P_{c1}$ represents the first critical pressure, below which $T_c$ displays a slight change. $P_{c2}$ stands for the second critical pressure, where the superconducting-insulation transition occurs. (b-e) The mapping information of temperature- and pressure-dependent resistance (see the color scale) for the samples, along with the phase diagrams established by the experimental results. (f) Plots of doping concentration ($x$) in $Ba_{1-x}K_xBiO_3$ and hole doping concentration ($p$) per Cu in $Bi_2Sr_2CaCu_2O_{8+\delta}$ versus their critical pressure ($P_{c2}$) for the SC-I phase transition, respectively. The diamonds are the data from $Ba_{1-x}K_xBiO_3$, while the circles are the data from $Bi_2Sr_2CaCu_2O_{8+\delta}$.

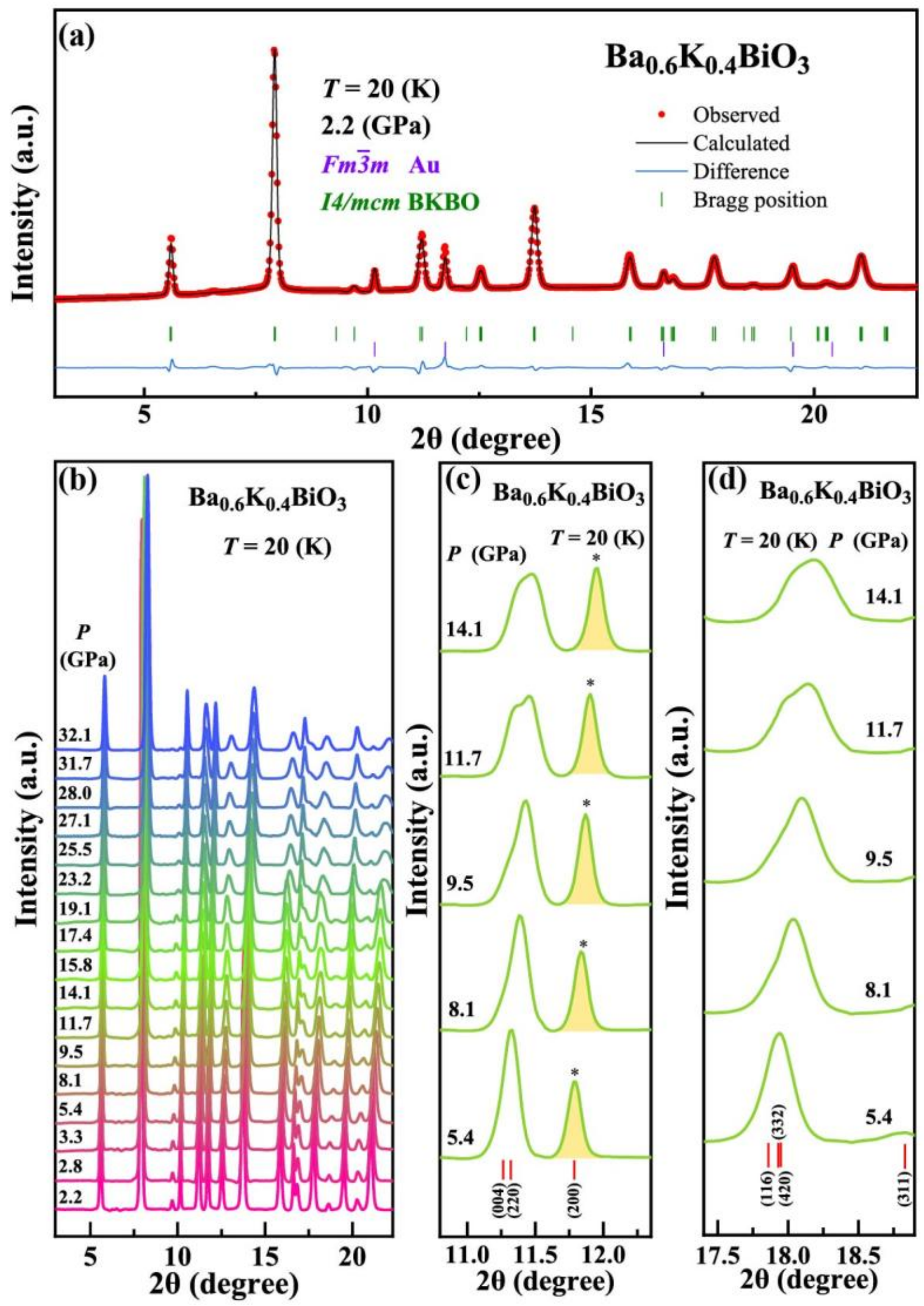

Figure 4 Structure information of the $Ba_{0.6}K_{0.4}BiO_3$ sample measured at 20 K and various pressure conditions. (a) Typical Rietveld structural refinements for the sample subjected to 2.2 GPa and 20 K, displaying that the sample crystalizes in a tetragonal unit cell with space group *I4/mcm*. (b) Selected X-ray diffraction patterns collected at 20 K and different pressures. (c)-(d) Enlarge views of low-temperature XRD patterns collected at different pressures, showing the splits of peaks of the tetragonal (T) phase. The colored peaks in the patterns correspond to elemental Au, which serves as the pressure marker in our study.

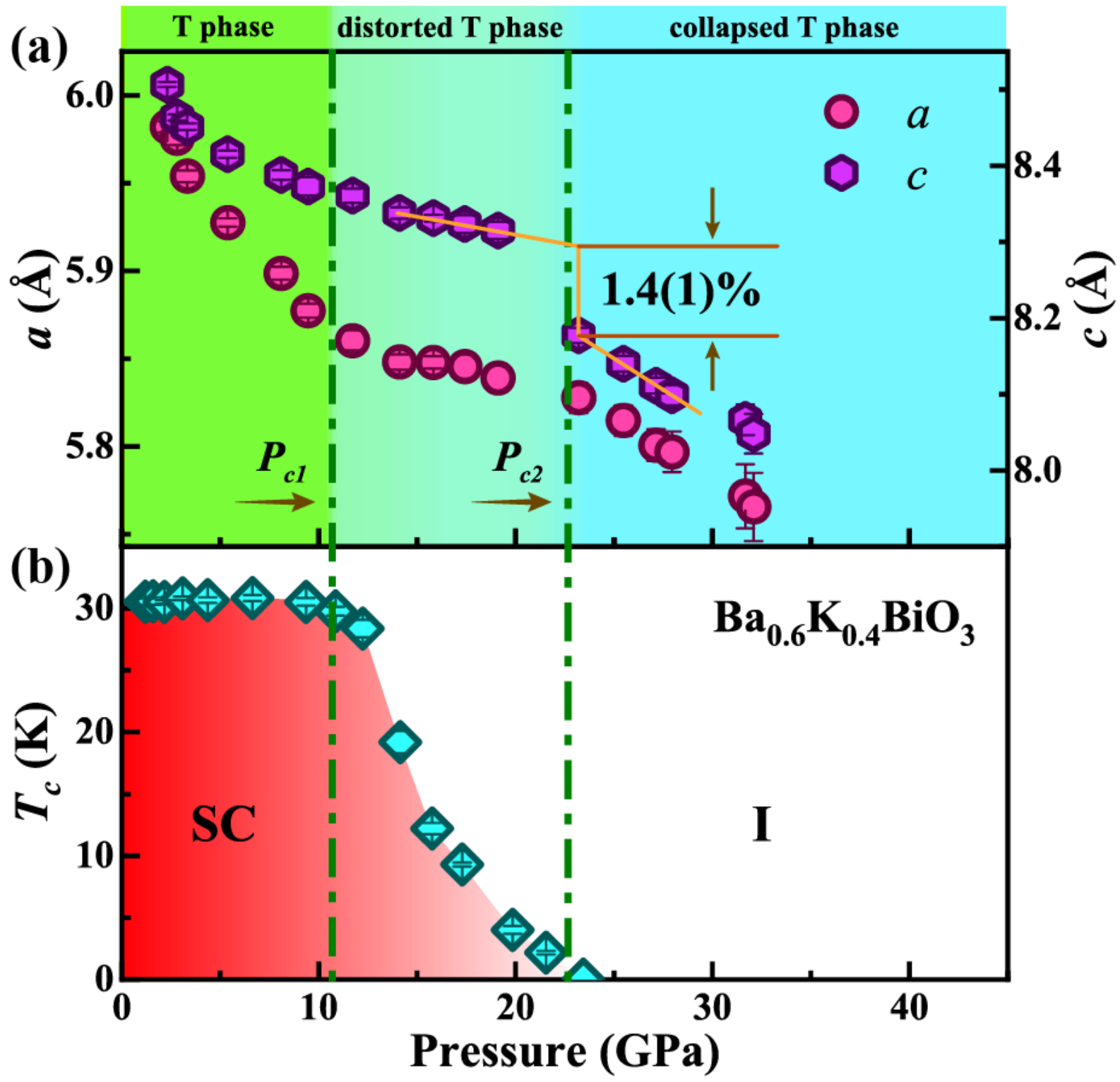

Figure 5 Pressure dependence of low-temperature (at 20 K) lattice parameters and superconducting transition temperature ($T_c$) for the $Ba_{0.6}K_{0.4}BiO_3$ sample. (a) Plots of lattice parameter *a* and *c* versus pressure obtained at 20 K. Below $P_{c1}$, the lattice parameter *a* and *c* of the T phase show a monotonous decrease with increasing pressure. In the pressure range between $P_{c1}$ and $P_{c2}$, the lattice parameter *a* and *c* of the T phase shrink anionotropic along the *a* and *c* axes. Beyond $P_{c2}$, the T phase collapses along the *c* axis. (b) $T_c$ as a function of pressure, showing the three distinct regions: $T_c$ exhibits small variation below $P_{c1}$ (left), a dramatic monotonous decrease between $P_{c1}$ and $P_{c2}$ (middle), and the appearance of the insulating state at pressure above $P_{c2}$ (right).